# A model for the compositions of non-stoichiometric intermediate phases formed by diffusion reactions, and its application to Nb$_3$Sn superconductors

X. Xu*, M. D. Sumption

Department of Materials Science and Engineering, the Ohio State University, Columbus, OH 43210 USA

*Electronic mail: xu.452@osu.edu

*In this work we explore the compositions of non-stoichiometric intermediate phases formed by diffusion reactions: a mathematical framework is developed and tested against the specific case of Nb$_3$Sn superconductors. In the first part, the governing equations for the bulk diffusion and inter-phase interface reactions during the growth of a compound are derived, numerical solutions to which give both the composition profile and growth rate of the compound layer. The analytic solutions are obtained with certain approximations made. In the second part, we explain an effect that the composition characteristics of compounds can be quite different depending on whether it is the bulk diffusion or grain boundary diffusion that dominates in the compounds, and that "frozen" bulk diffusion leads to unique composition characteristics quite distinct from equilibrium expectations; then the model is modified for the case of grain boundary diffusion. Finally, we apply this model to the Nb$_3$Sn superconductors and propose the approaches to control their compositions.*

## Introduction

Intermediate phases with finite composition ranges represent a large class of materials, and their compositions may influence their performance in application, as demonstrated in a variety of materials, such as electrical conductivity of oxides (e.g., TiO$_{2-y}$ [1]), electromagnetic properties of superconductors (e.g., Nb$_3$Sn and YBa$_2$Cu$_3$O$_{7-y}$ [2]), and mechanical properties of some intermetallics (e.g., Ni-Al$_{0.4-0.55}$ [3]), etc. For instance, the superconducting Nb$_3$Sn phase, which finds significant applications in the construction of 12-20 T magnets[4,5], has a composition range of ~17-26 Sn at.%[6,7], and its superconducting transition temperature and magnetic field decrease dramatically as Sn content drops[4,7-8]. The Nb$_3$Sn phase, which is generally fabricated from Cu-Sn and Nb precursors through reactive diffusion processes, is always found to be Sn-poor (e.g., 21-24 at.%[4,8-9]), making composition control one of the primary concerns in Nb$_3$Sn development since 1970s[10]. Although a large number of previous experiments (e.g., [4,8-11]) have uncovered some factors that influence the Sn content, it is still a puzzle what fundamentally determines the Nb$_3$Sn composition. This work aims to fill that gap. Here it is worth mentioning that the composition interval of a compound layer does not necessarily coincide with its equilibrium phase field ranges – the former can be narrower (e.g., the Nb$_3$Sn example above) if the inter-phase interface reaction rates are slow relative to the diffusion rate across the compound, which results in discontinuities in chemical potentials at the interfaces.

There have been numerous studies regarding diffusion reaction processes, most of which have focused on layer growth kinetics (e.g.,[12-16]), compound formation and instability (e.g.,[14-16]), phase diagram determination (e.g.,[17]), and interdiffusion coefficient measurement (e.g.,[18]), while a systematic model exploring how to control compound composition is still lacking. We find it indeed possible to modify the model developed by Gosele and Tu[13] for deriving the layer growth kinetics of compounds to calculate their compositions; however, certain assumptions (e.g., steady-state diffusion and first-order interface reaction rates) that the model was based on may limit the accuracy of the composition results. In this work, we aim to develop a rigorous, systematic mathematical framework for the compositions of intermediate phases.

**Results**

Let us consider that an $A_nB$ compound with a B content range of $X_l$ - $X_u$ ($X_l < X_u < 1$) is formed in a system of M-B/A, where M is a third element that does not dissolve in $A_nB$ lattice[19], and the solubility of B in A is negligible[6]. This is the case we see for the $Nb_3Sn$ example above (for which A stands for Nb, B for Sn, and M for Cu), but the work below can be modified for other cases. An isothermal cross section of such an M-A-B phase diagram at a certain temperature is shown in Fig. 1. The use of the third element M is to decrease the chemical potential of B, so that unwanted high-B A-B compounds (e.g., $NbSn_2$ and $Nb_6Sn_5$ [6]) that would form in the B/A binary system can be avoided. Similar to the Cu-Nb-Sn system, let us assume B is the primary diffusing species in the $A_nB$ layer[20] and that the diffusivity of B in M-B is high[10]. A schematic of the M-B/$A_nB$/A system for a planar geometry is shown in Fig. 2, but the model can be modified for cylindrical and spherical geometries. Let us denote the M-B/$A_nB$ and $A_nB$/A inter-phase interfaces as I and II, respectively, and the concentrations, chemical potentials, activities, and diffusion fluxes of B in the $A_nB$ layers near interfaces I and II as $X_I$, $\mu_I$, $a_I$, $J_I$, and $X_{II}$, $\mu_{II}$, $a_{II}$, $J_{II}$, respectively. Let us also denote the $X_B$s, $\mu_B$s, $a_B$s of M-B source and A-$X_l$ B as $X_s$, $\mu_s$, $a_s$, and $X_l$, $\mu_l$, $a_l$, respectively.

In this work let us assume the diffusivity of B in $A_nB$, $D$, and the molar volume of $A_nB$, $V_m$, do not vary with $X_B$, in which case the continuity equation in the $A_nB$ layer is given by:

$$\frac{\partial X_B}{\partial t} = D \frac{\partial^2 X_B}{\partial x^2} \qquad (1)$$

According to mass conservation, in a unit time the amount of B transferring across the interface I should equal to that diffusing into $A_nB$ from the interface I, and the amount arriving at the interface II should equal to that transferring across it, i.e., $dn/dt|_I = J_I \cdot A_I$, and $dn/dt|_{II} = J_{II} \cdot A_{II}$, where $A_I$ and $A_{II}$ are the areas of the interfaces I and II, respectively. The molar transport rate $dn/dt$ across an interface equals to $r \cdot A_{int} \cdot \exp(-Q/RT) \cdot [1-\exp(-\Delta\mu/RT)]$, where $r$ is the transfer rate constant for this interface with the unit of mol/(m²·s), $A_{int}$ is the interface area, $Q$ is the energy barrier, $R$ is the gas constant, $T$ is the temperature in K, and $\Delta\mu$ is the driving force for atom transfer. For the interface I, $\Delta\mu|_I = \mu_s - \mu_I$. For the interface II, $\Delta\mu|_{II} = \mu_{II} - \mu_l$, because $\mu_B(A) = \mu_B(A \text{-} X_l B)$. With $J_B = -(D/V_m) \cdot (\partial X_B / \partial x)$, we have:

$$r_I \exp\left(-\frac{Q_I}{RT}\right)\left[1 - \exp\left(-\frac{\mu_s - \mu_I}{RT}\right)\right] = -\frac{D}{V_m}\frac{\partial X_B}{\partial x}\Big|_I \qquad (2)$$

$$r_{II} \exp\left(-\frac{Q_{II}}{RT}\right)\left[1 - \exp\left(-\frac{\mu_{II} - \mu_l}{RT}\right)\right] = -\frac{D}{V_m}\frac{\partial X_B}{\partial x}\Big|_{II} \qquad (3)$$

Eqs. (2) and (3) are the boundary conditions for Eq. (1). Note that $X_s$ drops with annealing time as B in M-B is used for $A_nB$ growth, so $\mu_s$ drops with $t$:

$$\frac{d\mu_s}{dt} = \frac{d\mu_s}{dX_s}\frac{dX_s}{dt} = -\frac{d\mu_s}{dX_s}\frac{n_M}{\left(n_M + n_{B0} - \int_t dn/dt|_I\right)^2}\frac{dn}{dt}\Big|_I \qquad (4)$$

where $n_M$ and $n_{B0}$ are the moles of M and B in the M-B precursor. For those systems without the third element, $\mu_s$ is constant, and Eq. (4) is not needed. In addition, since the B atoms diffusing to the interface II are used to form new $A_nB$ layers, we have:

$$\frac{dl}{dt} = \frac{J_{II}V_m}{X_{II}} = -\frac{D}{X_{II}}\frac{\partial X_B}{\partial x}\Big|_{II} \qquad (5)$$

Eqs. (1)-(5) are the governing equations for the system set up above, solutions to which give both the $X_B(x, t)$ and the $l(t)$ of a growing $A_nB$ layer. It should be noted that for the systems with large volume expansion associated with transformation from A to $A_nB$, stress effects need to be considered[21].

To simplify Eqs. (2) and (3), we notice that $1 - \exp[-(\mu_s - \mu_I)/RT] = 1 - a_I/a_s$, since $\mu_s - \mu_I = RT\ln(a_s/a_I)$; similarly, $1 - \exp[-(\mu_{II} - \mu_l)/RT] = 1 - a_l/a_{II}$. Let us also denote $D/[V_m \cdot r_I \cdot \exp(-Q_I/RT)]$ as $\varphi_I$, and $D/[V_m \cdot r_{II} \cdot \exp(-Q_{II}/RT)]$ as $\varphi_{II}$: clearly $\varphi_I$ and $\varphi_{II}$ represent the ratios of diffusion rate over interface reaction rates, and have a unit of meter. Then Eqs. (2) and (3) can be respectively written as:

$$1 - \frac{a_I}{a_s} = -\varphi_I \frac{\partial X_B}{\partial x}\big|_I \qquad (6)$$

$$1 - \frac{a_I}{a_{II}} = -\varphi_{II} \frac{\partial X_B}{\partial x}\big|_{II} \qquad (7)$$

For now, let us consider two extreme cases.

First, for the case that the interface reaction rates are much higher than the diffusion rate across the $A_nB$ layer (i.e., diffusion-rate limited), $\varphi_I$ and $\varphi_{II}$ are near zero; according to Eqs. (2)-(3), $\mu_B$s are continuous at both interfaces. Suppose $\mu_s$ and the position of interface I, $x_I$, are both constant with time, then $X_I$ is also constant, and the solutions to Eqs. (1) and (5) are respectively $X_B(x, t) = X_I-(X_I-X_l)\cdot\mathrm{erf}\{(x-x_I)/[2\sqrt{(Dt)}]\}/\mathrm{erf}(k/2)$ and $l=k\sqrt{(Dt)}$ for the $A_nB$ layer, where $k$ can be numerically solved from $k\cdot\exp(k^2/4)\cdot\mathrm{erf}(k/2) = 2/\sqrt{\pi}\cdot(X_I-X_l)/X_l$. For instance, for $X_I = 0.26$ and $X_l = 0.17$, $k=0.953$. On the other hand, if the interface reaction rates are much lower than the diffusion rate across $A_nB$ (e.g., as the $A_nB$ layer is thin), $\varphi_I$ and $\varphi_{II}$ are large; according to Eqs. (2)-(3), $X_B$ and $J_B$ are nearly constant in the entire $A_nB$ layer. Thus, $(1-a_B/a_s)/\varphi_I = (1-a_l/a_B)/\varphi_{II}$, from which $a_B$ can be calculated. Integration of Eq. (5) gives: $l \propto t$, and the pre-factor depends on the interface reaction rates.

For a general case between these two extremes, the equations can only be solved with the $\mu(X)$ relations of M-B and $A_nB$ provided. Next, let us consider a compound with a narrow composition range, so that as a Taylor series expansion is performed around $X_l$ for its $a(X_B)$ curve, high-rank terms can be neglected because $|X-X_l| \leq (X_u-X_l)$ is small; that is, $a_X \approx a_l + \kappa(X-X_l)$, where $\kappa$ is the linear coefficient of the $a(X)$ curve. Given the complex boundary conditions for Eq. (1), to obtain the analytic solutions we introduce a second approximation if the $A_nB$ composition range is narrow: the $X(x)$ profile of the $A_nB$ layer is linear so that at a certain time $J$ is constant with $x$, such that $-(\partial X_B/\partial x)_I \approx -(\partial X_B/\partial x)_{II} \approx (X_I-X_{II})/l$. With these two approximations, we can solve Eqs. (6)-(7) and obtain that:

$$a_{II} = \frac{\sqrt{(\varphi_I a_s + \kappa l - \varphi_{II} a_s)^2 + 4\varphi_{II} a_I (\varphi_I a_s + \kappa l)} - (\varphi_I a_s + \kappa l - \varphi_{II} a_s)}{2\varphi_{II}} = \frac{2a_I}{1-\eta + \sqrt{(1-\eta)^2 + 4\eta\,a_I/a_s}} \quad (8)$$

where $\eta = \varphi_{II} a_s/(\varphi_I a_s + \kappa l)$. Then $a_I$ can be calculated from $a_{II}$, and $X_I$ and $X_{II}$ can be calculated from $a_I$ and $a_{II}$ using $X = X_l + (a_X - a_l)/\kappa$.

To verify the results, the equations are solved for a hypothetical system analytically and numerically, with and without the assumption that $X(x)$ is linear, respectively. The obtained composition profiles are shown in Fig. 3 (a). For simplicity, $\mu_s$ of the system is set as $\mu_B(A\text{-}X_u\,B)$ and is constant (for Nb$_3$Sn systems, this means that Nb$_6$Sn$_5$ serves as Sn source), and the other parameters are specified in the figure. The difference between the analytic and numerical solutions is $<0.1\%$, showing that the approximation of linear $X(x)$ is good if the composition range is small (2 at.% in this case). The $l(t)$ result (where $t$ is the annealing time after the incubation period) from the numerical calculations is shown in Fig. 3 (b). While the analytic $l(t)$ solution is complicated, some $l(t)$ relations with simple forms can be used as approximations. The most widely used $l(t)$ relation for the case of constant $\mu_s$ is $l=bt^m$, in which $m=1$ for reaction-rate limited and $m=0.5$ for diffusion-rate limited; however, a defect with this relation is that as $l$ increases from zero, it may shift from reaction-rate limited to diffusion-rate limited, so $m$ may vary with $t$. Here a new relation $l=q[\sqrt{(t+\tau)}-\sqrt{\tau}]$ (where $q$ is a growth constant and $\tau$ is a characteristic time) is proposed. Such a relation is consistent with $l^2/v_1 + l/v_2 = t$ (where $v_1$ and $v_2$ are constants related to diffusion rate and interface reaction rates, respectively) proposed by previous studies[13,14]. This relation overcomes the above problem because as $t \ll \tau$, $l = [q/(2\tau)] \cdot t$ and as $t >> \tau$, $l=q\sqrt{t}$. As can be seen from Fig. 3 (b), a better fit to the numerical $l(t)$ curve in the whole range is achieved by $l=q(\sqrt{(t+\tau)}-\sqrt{\tau})$.

## Discussion

Before discussing the application of this model to a specific material system, it must be pointed out that all of the analysis and calculations above are for the case that B diffuses through $A_nB$ bulk. In such a case, for an M-B/$A_nB$/A system, as $\mu_s$ drops with the growth of $A_nB$ layer, $X_B(x)$ of $A_nB$ should decrease with $\mu_s$, because $\mu_s \geq \mu_I \geq \mu_{II} \geq \mu_l$. Finally, one of two cases will occur: either $\mu_s$ drops to $\mu_l$ (if A is in excess) so the system ends up with the equilibrium among A, A-$X_l$ B, and M-$X_l$ B (as shown by the shaded region in the isothermal M-A-B phase diagram in Fig. 1), or A is consumed up and $A_nB$ gets homogenized with time and finally $\mu_B(A_nB)=\mu_B(\text{M-} B)$ (as shown by the dashed line in Fig. 1). In either case, $A_nB$ eventually reaches homogeneity.

However, we find that the composition could be different for a compound in which the bulk diffusion is low while grain boundary diffusion dominates. One such example is $Nb_3Sn$, the composition of which displays some extraordinary features. As an illustration, the $X_{Sn}$s of a Cu-Sn/$Nb_3Sn$/Nb diffusion reaction couple after various annealing times are shown in Fig. 4. Clearly, as the $X_{Sn}$ (and $\mu_{Sn}$) of Cu-Sn drop with time, the $X_{Sn}$s of $Nb_3Sn$ do not drop accordingly; instead, they more or less remain constant with time. In addition, from 320 hours to 600 hours, although Nb has been fully consumed, the $X_{Sn}$ of $Nb_3Sn$ does not homogenize (i.e., the $X_{Sn}$ gradient does not decrease) with time. In many other studies on Cu-Sn/Nb systems with Nb in excess (e.g., [4,8-9]), even after extended annealing times after the $Nb_3Sn$ layers have finished growing (which indicates that the Sn sources have been depleted, i.e., $\mu_{Sn}$s have dropped to $\mu_l$), $X_{Sn}$s of $Nb_3Sn$ remain high above $X_l$, without dropping with annealing time.

The reason for these peculiarities is that grain boundary diffusion in $Nb_3Sn$ dominates due to extremely low bulk diffusivity (e.g., lower than $10^{-23}$ $m^2$/s at 650 °C)[20,22,23] and small $Nb_3Sn$ grain size (~100 nm). In this case, our model and equilibrium-state analysis apply only to the diffusion zones (i.e., the grain boundaries and the inter-phase interfaces) instead of the bulk. To clarify this point more clearly, a schematic of the diffusion reaction process is shown in Fig. 5. At time $t_1$, at the $A_nB$/A interface, high-B $A_nB$ ($L_2$ layer) reacts with A ($L_3$ layer) to form some new $A_nB$ cells, leaving B vacancies (noted as $V_Bs$) in $L_2$ layer (time $t_2$). If bulk diffusivity is high, $V_Bs$ simply diffuse through bulk (e.g., from $L_2$ to $L_1$, as shown by grey dotted arrows) to the B source. If bulk diffusion is frozen, the $V_Bs$ diffuse first along $A_nB$/A inter-phase interface (as shown by green solid arrows), and then along $A_nB$ grain boundaries to the B source. This process continues until this $L_3$ layer entirely becomes $A_nB$ (time $t_3$), so the reaction frontier moves ahead to $L_3$/$L_4$, while the $L_2$/$L_3$ interface now becomes an inter-plane inside $A_nB$ lattice. If bulk diffusion is frozen, the $V_Bs$ in the $L_2$ layer that have not diffused to B source will be frozen in this layer forever, and will perhaps transform to A-on-B anti-site defects later (e.g., for $Nb_3Sn$, Nb-on-Sn anti-sites are more stable than Sn vacancies[24]). Since these point defects determine the $A_nB$ composition, the $X_B$ in this $L_2$ layer cannot change anymore regardless of $\mu_B$ variations in grain boundaries. That is to say, $X_B$ of any point is just the $X_{II}$ of the moment when the reaction frontier sweeps across this point, i.e., the $X_B(x)$ of an $A_nB$ layer is simply an accumulation of $X_{II}$s with $l$ increase. Returning to Fig. 3 (a), the dashed lines display the evolution of $X_B(x)$ with $l$ increase for bulk diffusion, while that for grain boundary diffusion is shown by the solid lines. Since the energy dispersive spectroscopy (EDS) attached to scanning electron microscopes (SEM) that is used to measure the compositions typically has a spatial resolution of 0.5-2 μm, and thus mainly reflects the bulk composition, the composition characteristics of $Nb_3Sn$ layers as described above can be explained. It should be noted that knowledge of the difference between bulk diffusion and grain boundary diffusion is important in controlling the final composition of a compound. For instance, if bulk diffusivity is high, one method to form high-B $A_nB$ is increasing the starting B/A ratio so that after long annealing time for homogenization subsequent to the full consumption of A, $\mu_B$(M-B)=$\mu_B$(A-$X_u$ B). However, our experiments demonstrate that for compounds with low bulk diffusivity (e.g., $Nb_3Sn$), such an approach does not work; instead, controlling the $X_{II}$s while the compounds are growing is the only way. For those compounds with low but non-negligible bulk diffusivities, their compositions would be between these two extremes.

Then what determines the bulk composition as grain boundary diffusion dominates? From Fig. 5, it can be clearly seen that there is a competition deciding the $V_B$ fraction in the frontier $A_nB$ layer: at $t_2$ the reaction across the $A_nB$/A interface produces $V_B$s in $L_2$ layer, while the diffusion of B along $A_nB$ GBs and $A_nB$/A interface fills these $V_B$s. Thus, if the diffusion rate is slow relative to the reaction rate at interface II (i.e., $\varphi_{II}$ is low), a high fraction of $V_B$s would be left behind as the interface II moves ahead, causing low B content; if, on the other hand, the diffusion rate is high relative to the reaction rate at interface II, the $A_nB$ layer has enough time to get homogenized with the B source, causing low $X_B$ gradient. In this case, the $\mu_B$ of B source and the reaction rate at interface I together set a upper limit for $\mu_B$ of $A_nB$.

Next, we will modify the above model for the case of grain boundary diffusion for quantitative analysis. As pointed out earlier, the chemical potentials of grain boundaries can change with $\mu_s$ and $l$, while those of the bulk cannot. In such a case, $\mu_I$ and $\mu_{II}$ (suppose the diffusivities along the inter-phase interfaces are large) can still be calculated using our model, provided that the $\mu(X)$ relation and $D$ of the $A_nB$ grain boundary (instead of the bulk) are used in all of the equations, and that $\varphi_I$ and $\varphi_{II}$ are multiplied by a factor of $\sum A_{GB}/A_{int}$ (where $\sum A_{GB}$ is the sum of the cross section areas of the grain boundaries projected to the inter-phase interfaces), because B diffuses only along grain boundaries in $A_nB$ while reactions occur at the entire interfaces. Approximately, $\sum A_{GB}/A_{int} \approx [1-d^2/(d+w)^2] \approx 2w/d$ (where $w$ is the $A_nB$ grain boundary width, and $d$ is the grain size). Apparently, grain growth with annealing time reduces the diffusion rate. According to Eq. (8), $a_{II}$ is determined by $\eta$ and $a_s$, and increases with them, as shown by Fig. 6. Since $\eta=\varphi_{II}a_s/(\varphi_I a_s+\kappa l)= 1/[\varphi_I/\varphi_{II}+\kappa l/(\varphi_{II}a_s)]$, clearly $\eta$ decreases as $\varphi_I/\varphi_{II}$ and $l$ increase, and the influence of $l$ (which reflects the $X_{II}$-$x$ gradient) is mitigated as $\varphi_{II}a_s$ increases. Thus, to improve $X_{II}$ of $A_nB$ at $l=0$, one should increase $\mu_s$ and the reaction rate at interface I, and reduce the reaction rate at interface II; meanwhile, to reduce $X_{II}(x)$ gradient, one should increase $\varphi_{II}$ (which means improving the diffusion rate or reducing the reaction rate at interface II) and $a_s$. Apparently, these quantitative conclusions are consistent with the above qualitative analysis.

Next let us compare this model with the example of $Nb_3Sn$. It has been well established from experimental work that there are mainly two factors that can significantly influence the Sn content of $Nb_3Sn$ in a $Cu$-$Sn/Nb_3Sn/Nb$ diffusion reaction couple: heat treatment temperature and Cu-Sn source. The heat treatment temperature can simultaneously influence multiple factors of Eq. (8), such as $a_s$, $D$, and reaction rates at both interfaces, etc. Thus, the explanation of the influence of temperature on Sn contents using this theory requires knowledge of the quantitative variations of these factors with temperature. For the other factor, Cu-Sn source, the diffusion reaction couples can be classified into two types based on the Cu-Sn source: the type I uses bronze (with Sn content in Cu-Sn typically below 9 at.%) as Sn source, and the type II uses high-Sn Cu-Sn (e.g., Cu-25 at.% Sn). Previous measurements[4,8-10,25] show that both types of samples have Sn contents above 24 at.% for the $Nb_3Sn$ layer next to the Cu-Sn source; however, they have distinct Sn content gradients as the $Nb_3Sn$ layers grow thicker: the type I generally has Sn content gradients above 3 at.%/μm[25], while those of the type II are below 0.5 at.%/μm[4,8-9]. Such a difference in the Sn content gradients leads to distinct grain morphologies and superconducting properties. The different $X_{Sn}$ gradients in the two types of samples with different Cu-Sn sources can be easily explained by our theory above: according to Eq. (8), increased $\mu_s$ can decrease $X_{Sn}$ gradients. It may also need further investigation regarding whether Cu-Sn source can also influence diffusion rates in Nb$_3$Sn layer (e.g., via thermodynamic factor), because greater $D$ leads to greater $\varphi_{II}$, which helps decreasing $X_{Sn}$ gradients. As to the phenomenon that different wires have similar $X_{Sn}$ in the Nb$_3$Sn layer next to the Cu-Sn source, the relation between $\mu_{Sn}$(Cu-Sn) and $\mu_{Sn}$(Nb-$X_{Sn}$ Sn) is required. The Cu-Sn system has been well studied, and the phase diagram calculated by the CALPHAD technique using the thermodynamic parameters given by Ref. 26 is well consistent with the experimentally measured diagram[27]. Thus, the parameters from Ref. 26 are used to calculate $\mu_{Sn}$ of Cu-Sn, which is shown in Fig. 7. On the other hand, although thermodynamic data of Nb-Sn system were proposed by Refs. 26 and 28, in these studies Nb$_3$Sn was treated as a line compound. However, some information about $\mu_{Sn}$ of Nb$_3$Sn can be inferred from its relation with $\mu_{Sn}$ of Cu-Sn: since Cu-7 at.% Sn leads to the formation of Nb-24 at.% Sn near the Cu-Sn source[25], we have $\mu_{Sn}$(Cu-7 at.% Sn) $\geq \mu_{Sn}$(Nb-24 at.% Sn). Thus, the expected approximate $\mu_{Sn}$(Nb-$X_{Sn}$ Sn) curve in a power function is shown in Fig. 7. Besides, we can also infer that the Sn transfer rate at the Cu-Sn/Nb$_3$Sn interface must be much faster than that at the Nb$_3$Sn/Nb interface, so $\mu_{Sn}$ discontinuity across the interface I is small. These explain why low-Sn Cu-Sn can lead to the formation of high-Sn Nb$_3$Sn. It is worth mentioning that from Fig. 7, it is clear that the Taylor series for the true $a(X)$ relation of Nb$_3$Sn have more high-rank terms than $a(X) \approx a_l + \kappa(X-X_l)$; however, our numerical calculations show that adding high-rank terms to the $a(X)$ relation does not lead to different conclusions regarding the influences of $a_s$, $\varphi_I$, $\varphi_{II}$, and $l$ on $X_{II}$. Thus, the above qualitative and quantitative analysis still applies.

In summary, a mathematical framework is developed to describe the compositions and layer growth rates of non-stoichiometric intermediate phases formed by diffusion reactions. The governing equations are derived and analytic solutions are given for compounds with narrow composition ranges under certain approximations. We also modify our model for compounds in which bulk diffusion is frozen, of which the bulk is not in equilibrium with the rest of the system. Based on this model, the factors that fundamentally determine the compositions of non-stoichiometric compounds formed by diffusion reactions are found and approaches to control the compositions are proposed.

## Methods

For the Cu-Sn/Nb$_3$Sn/Nb diffusion reaction couples that were used for Sn content measurements (the results of which are shown in Fig. 4), the initial composition of the precursor Cu-Sn alloy was Cu-12 at.% Sn. The samples were reacted at 650 °C for 65 h, 130 h, 320 h, and 600 h. Then the samples were polished to 0.05 μm and the compositions were measured using an EDS system attached to an SEM. An accelerating voltage of 15 kV was used for the quantitative line scans. A standard Nb-25 at.% Sn bulk sample provided by Dr. Goldacker from Karlsruhe Institute of Technology was used for calibrating the Sn content of the samples. The standard deviation in the measurements was found to be about ± 0.5 at.%.

**Acknowledgements**

The authors thank S. Dregia and J. Morral for useful discussions, and X. Peng and Hyper Tech Research Inc. for providing Nb$_3$Sn samples for analysis. The work is funded by the US Department of Energy, Division of High Energy Physics, under an SBIR program.

**Author Contributions**

X.X initiated this study, developed the model, and wrote the manuscript. M.D.S. supported this work, discussed the results, and reviewed the manuscript.

**Additional information:**

**Figure Captions**

FIG. 1: Schematic of an isothermal cross section of the M-A-B ternary phase diagram. The shaded region shows the equilibria among M-$X_I$ B, A-$X_I$ B, and A phases, and the dashed line shows the equilibrium between M-B and $A_nB$ phases.

FIG. 2: (a) Schematic of the M-B/$A_nB$/A diffusion reaction system in the planar geometry, and (b) $X_B$ profiles of the system.

FIG. 3: (a) The calculated $X_B(x)$ profiles of the hypothetical system for the analytic and numerical solutions, with and without the assumption that $X_B(x)$ is linear, respectively. (b) The $l(t)$ results from the numerical calculations, with the fits of $l=q[\sqrt{(t+\tau)}-\sqrt{\tau}]$ and $l=bt^m$.

FIG. 4: The measured $X_{Sn}$s of a Cu-Sn/$Nb_3Sn$/Nb system after various annealing times at 650°C. The standard deviation in the Sn content measurements is around $\pm$ 0.5 at.%.

FIG. 5: A schematic of the diffusion reaction process for grain boundary diffusion.

FIG. 6: The variation of $a_{II}$ with $\eta$ and $a_s$, according to Eq. (8).

FIG. 7: The variation of $\mu_{Sn}$ with $X_{Sn}$ for Cu-Sn calculated based on thermodynamic data given in Ref. 26, and a rough, speculative $\mu_{Sn}(X_{Sn})$ relation for $Nb_3Sn$ sketched according to the phase formation relation between Cu-Sn and $Nb_3Sn$.

FIG 1

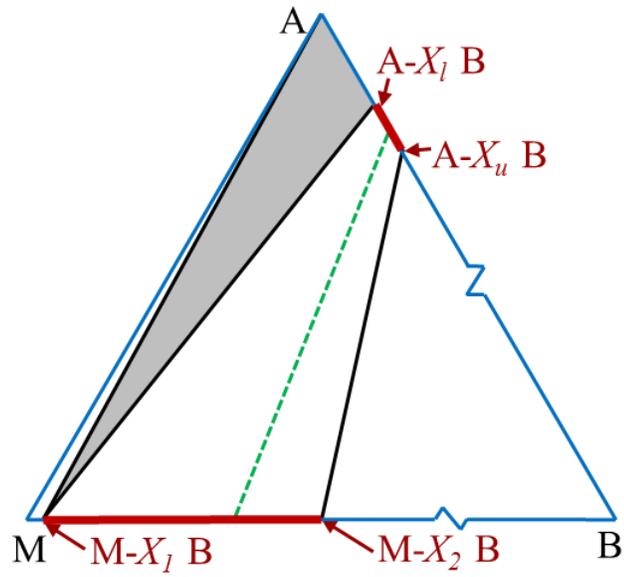

FIG 2

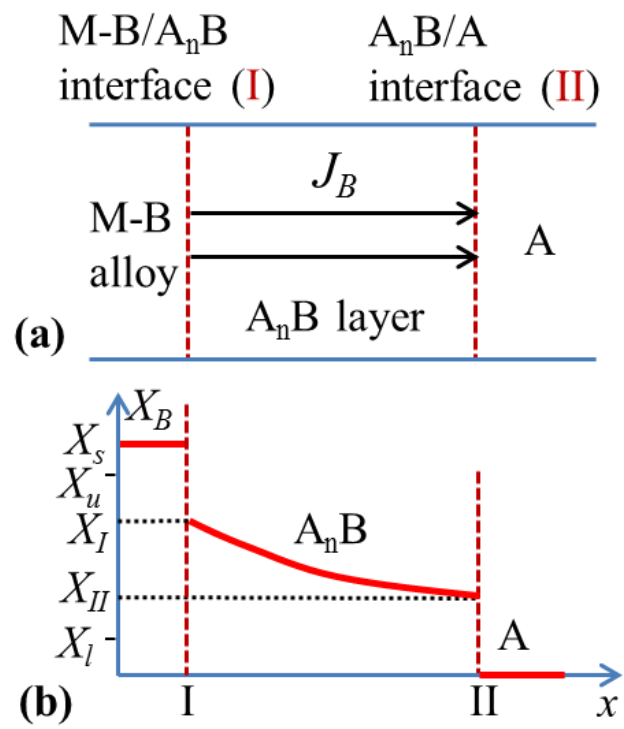

**(a)**

**(b)**

FIG 3

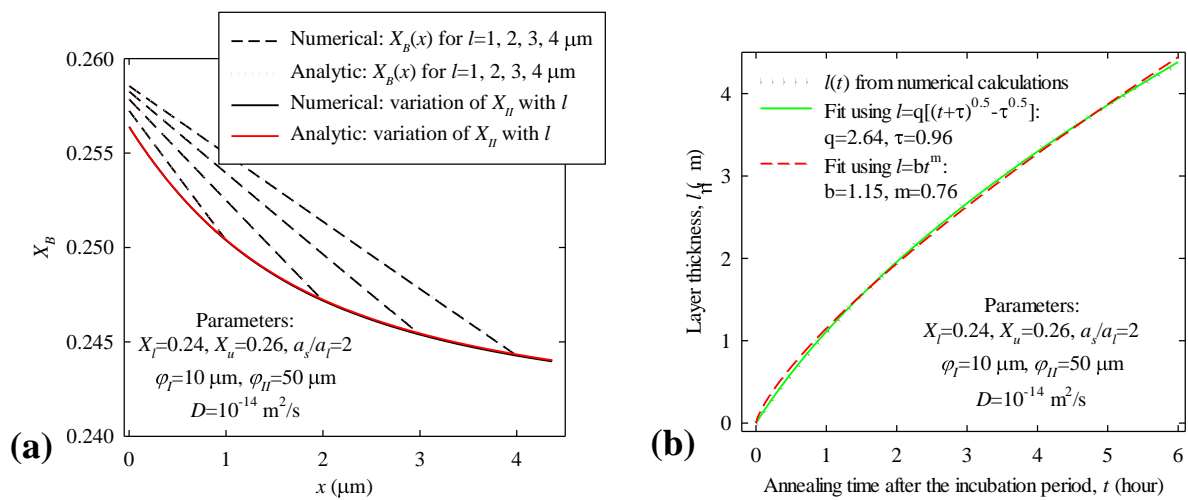

FIG 4

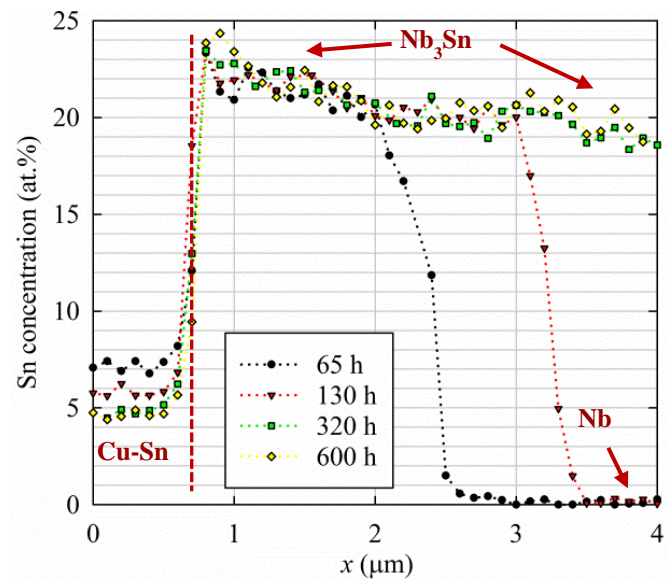

FIG 5

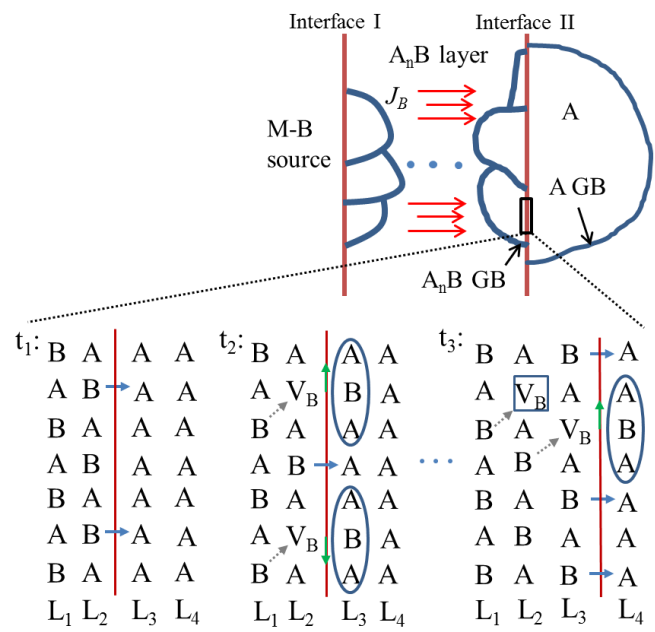

FIG 6

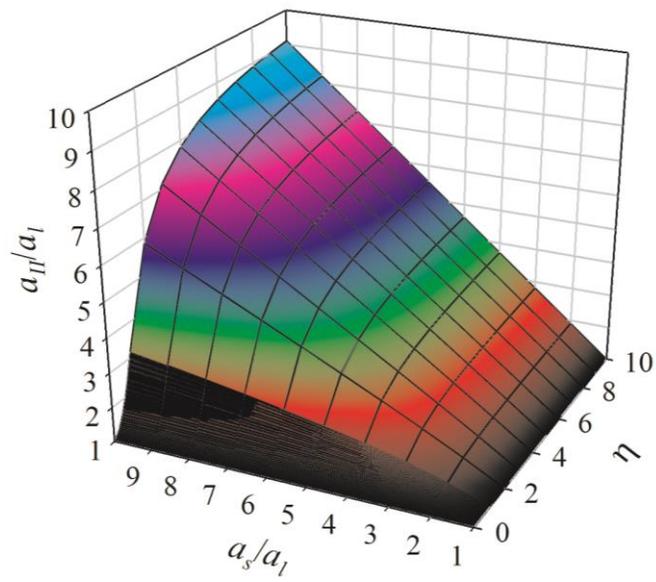

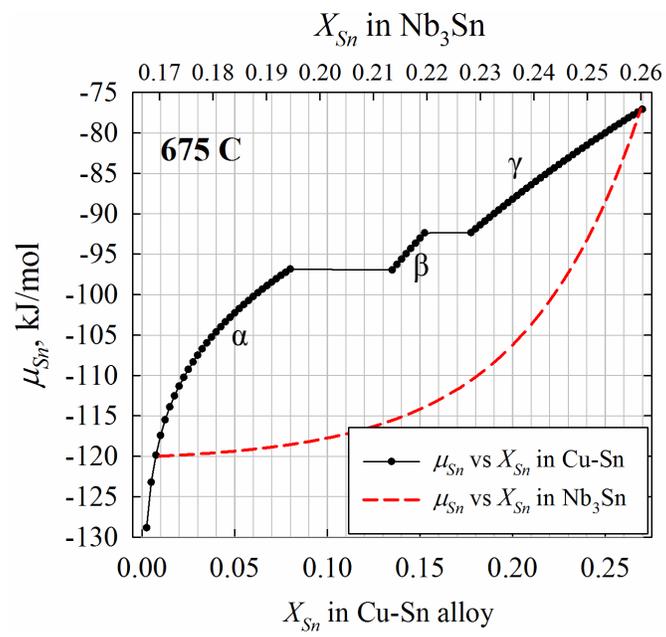